\documentclass[%
 reprint,
 amsmath,amssymb,
 aps,
]{revtex4-2}

\usepackage{graphicx}% Include figure files
\usepackage{dcolumn}% Align table columns on decimal point
\usepackage{bm}% bold math
\def\u{{\mathbf{u}}}

\begin{document}

\title{Kramers–Wannier Duality and Random Bond Ising Model}

\author{Chaoming Song}%
\email{c.song@miami.edu}
\affiliation{%
Department of Physics, University of Miami, Coral Gables FL, 33146 USA.
}%

\date{\today}% It is always \today, today,

\begin{abstract}
We present a new combinatorial approach to the Ising model incorporating arbitrary bond weights on planar graphs. In contrast to existing methodologies, the exact free energy is expressed as the determinant of a set of ordered and disordered operators defined on vertices and dual vertices respectively, thereby explicitly demonstrating the Kramers–Wannier duality. The implications of our derived formula for the random bond Ising model are further elucidated.
\end{abstract}

\maketitle

{\bf Introduction:} 
The well-established duality between order and disorder phases observed in the two-dimensional Ising model was initially exploited by Kramers and Wannier to pinpoint its criticality \cite{kramers1941statistics}, predating the celebrated solution of Onsager's free energy \cite{onsager1944crystal, kaufman1949crystal,kaufman1949crystalb}. Furthermore, Kadanoff and Ceva illustrate that the correlation function can be derived by contemplating the disorder operator defined on the dual graph \cite{kadanoff1971determination}, providing substantial insights into the underlying physics \cite{fradkin2017disorder}. However, the standard methodologies used for the computation of free energy—either algebraically or combinatorially \cite{mccoy1973two}—exhibit the Kramers-Wannier (KW) duality only in the final form after extensive calculations. Despite a tremendous volume of work over the last century devoted to the exact solution of the Ising model, a formula for calculating its free energy with manifest KW duality is still absent in the literature. This gap limits the utility of the exact free energy in broader contexts. For example, in the case of a random-bond Ising model (RBIM) employed for understanding spin glasses, deriving an explicit free energy remains challenging.

It is worth mentioning that the combinatorial approach, pioneered by Kac and Ward \cite{kac1952combinatorial}, provides an alternative pathway to derive Onsager's free energy. The Kac-Ward methodology hinges on the path-integral form on a planar graph $G = (V, E)$ with $n = |V|$ and $m = |E|$, which is based on an elegant identity
\begin{align}\label{eq:Fzeta}
\zeta_F(G,u)^{-1} \equiv \prod_{[p]} \left( 1- (-1)^{w(p)}u^{l(p)} \right) = Z_\textrm{ising}^{2} (1-u^2)^{m},
\end{align}
where $u = \tanh(\beta J)$ represents the coupling constant. This identity draws an analogy with Riemann's zeta function, where the Euler product is over all prime cycles, with $l(p)$ and $w(p)$ representing the length and winding number of the prime cycles $p$, respectively. The subscript $F$ highlights the fermionic character of the Ising model, which assigns a negative weight to odd numbers of windings. Initially conjectured by Feynman~\cite{sherman1960combinatorial,burgoyne1963remarks}, identity (\ref{eq:Fzeta}) was later formally proved by Sherman \cite{sherman1960combinatorial,sherman1962combinatorial,sherman1963addendum} and Burgoyne \cite{burgoyne1963remarks}. Based on Eq.~(\ref{eq:Fzeta}), Kac and Ward demonstrated that
\begin{equation}\label{eq:Fedge}
\zeta_F(G,u)^{-1} = \det (I_{2m} - u T_{KW}),
\end{equation}
where the Kac-Ward (KW) operator $T_{KW}$ is a $2m\times 2m$ matrix, further elucidated in the subsequent discussion. A variant of the combinatorial formulation, mapping the Ising model to the dimer model using Pfaffians, was developed by Green and Hurst \cite{hurst1960new} and later expanded by others \cite{potts1955combinatrial, kasteleyn1963dimer,montroll1963correlations,fisher1966dimer}. This approach essentially corresponds to a skew-symmetric version of Eq.~(\ref{eq:Fedge}) via a similarity transformation. More recently, a resurgence of the combinatorial approach \cite{mercat2001discrete,smirnov2007towards,smirnov2010conformal} focuses on the discrete version of the conformal invariance of the critical Ising model on planar graphs \cite{chelkak2017revisiting}.

The combinatorial approach can be seamlessly generalized to accommodate an arbitrary set of bond weights  $\u = { u_e| e\in E}$, thereby presenting a robust numerical tool for probing RBIM. However, Eq.~(\ref{eq:Fedge}) reveals little physical insight. For example, the KW duality indicates that $\zeta_F$ should remain invariant (up to some prefactor) under the transformation $u\to u^* = (1-u)/(1+u)$ on the dual graph $G^*$, i.e., $\zeta_F(G,u) \sim \zeta_F(G^*, u^*)$. Regrettably, the manifestation of the KW duality only emerges following the resolution of the determinant, a process that is impractical for numerous disordered systems. This hidden symmetry within Eq.~(\ref{eq:Fedge}) remains far from obvious and poses a significant challenge. It was only recently proven that Eq.~(\ref{eq:Fedge}) indeed satisfies KW duality under general conditions~\cite{cimasoni2012critical,cimasoni2015kac}. Consequently, despite its elegance, the combinatorial approach is primarily utilized as a numerical tool. To the best knowledge of the author, no explicit free energy formula demonstrating manifest KW duality for an arbitrary planar graph and weight set has yet been identified.

In this letter, we propose a new free-energy formula for the Ising model with arbitrary bond weights on planar graphs. Contrasting with existing methodologies, our formula is expressed as the determinant of the summation of local ordered and disordered operators, each defined on vertices $V$ and dual vertices $V^*$, thereby explicitly exhibiting the KW duality. In addition, it establishes a tangible connection with non-local ordered and disordered operators, offering insights into the nature of duality. We elucidate the implications of our formula in the context of RBIM.

\begin{figure*}
  \includegraphics[width=1\linewidth]{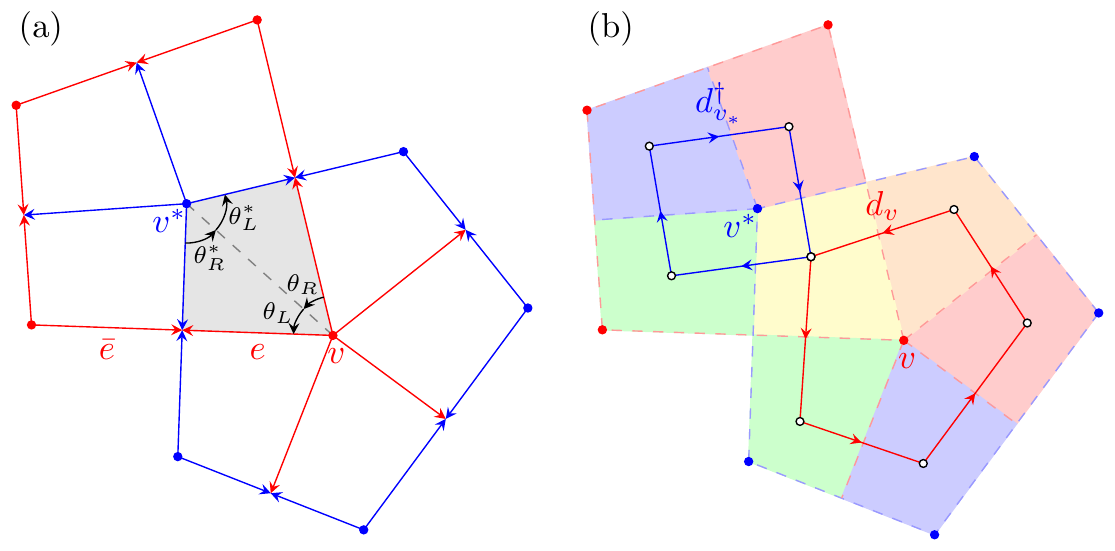}
  \caption{(a) The embedding of both $G$ and its dual $G^*$. The quadrilateral $q$ is delineated by a vertex $v$ and a neighboring dual vertex $v^*$, along with their respective edges. The relationships $\theta_L + \theta_R^* = \theta_R + \theta_L^* = \pi/2$ are satisfied. (b) The local order and disorder operators $d_v$ and $d_{v^*}^\dag$ for quadrilaterals. Each operator acts as a curl operator around the vertex $v$ and the dual $v^{*}$, respectively.
}
  \label{fig:embed}
\end{figure*}

{\bf Ihara zeta function:}  
To hint at the existence of the manifest dual formula, we initiate our discussion with a warm-up exercise by considering the bosonic counterpart of Eq.~(\ref{eq:Fzeta}),
\begin{align}\label{eq:Bzeta}
    \zeta_B(u)^{-1} = \prod_{[p]} \left( 1- u^{l(p)} \right).
\end{align}
This definition, known as the Ihara zeta function \cite{ihara1966discrete,sunada2006functions}, serves as a p-adic analogue of the Selberg zeta function that counts the number of closed geodesics on a hyperbolic surface. Analogous to Eq.~(\ref{eq:Fedge}), we express
\begin{equation}\label{eq:BEdge}
    \zeta_B(u)^{-1} = \det (I_{2m} - u T),
\end{equation}
where $T$ denotes the $2m\times 2m$ Hashimoto's edge adjacency operator \cite{hashimoto1989zeta} in analogy to the KW operator $K_{KW}$. Specifically, $T$ applies on the space of oriented edges $\{E,\bar E\}$, where an edge $\bar e \in \bar E$ symbolizes the directional inverse of a corresponding edge $e \in E$. The matrix $T_{e',e} = 1$ only if the oriented edge $e$ follows $e'$ backtracklessly, meaning that the terminal vertex of $e$ is the starting vertex of $e'$ and $e'\neq \bar e$. The equivalence of Eq.~(\ref{eq:Bzeta}) and Eq.~(\ref{eq:BEdge}) can be demonstrated directly by applying the logarithm and aligning the power expansion term by term.

For a regular graph of degree $q+1$, the Ihara zeta function displays self-duality under the transformation $u\to q/u$, which mirrors Riemann's functional equation. However, in similarity to its fermionic counterpart, Eq.~(\ref{eq:BEdge}) does not explicitly reveal this duality. Intriguingly, a second formula exists, as proposed in Ihara's original paper \cite{ihara1966discrete},
\begin{align}\label{eq:bass}
    \zeta_B(u)^{-1} = (1-u^2)^{m-n}\det (I_n -u A + u^2 Q),
\end{align}
where $A$ is the adjacency matrix, and $Q$ is the diagonal matrix with the degree diminished by one. Assuming the graph is regular, that is, $Q = qI_n$, it becomes evident that $\zeta_B^{-1}(G,u) \sim 
\zeta_B^{-1}(G,q/u)$. Consequently, Eq.~(\ref{eq:bass}) presents a manifestly dual formula for the bosonic zeta function. 

The derivation of Eq.~(\ref{eq:bass}) from Eq.~(\ref{eq:BEdge}) provides illuminating insights. Here, we present a streamlined approach based on the original proof by Bass~\cite{bass1992ihara,foata1999combinatorial,terras2010stroll}. We introduce the matrix $S = T + J$, where $J_{e',e} = \delta_{e',\bar e}$ and $S_{e',e}$ enumerates all successors $e'$ following $e$, including its inverse $\bar e$. A key observation arises from the factorization of matrix $S = Y^t X$, where $X$ and $Y$ are $n\times 2m$ matrices with $X_{v,e} = 1$ or $Y_{v,e} = 1$ if $v$ is the starting vertex or the terminal vertex of the orient edge $e$, respectively. Leveraging this factorization, we obtain  
\begin{align}
    \det (&I_m - u T)= \det \left((I_{2m} +u J)- u S \right) \notag \\
    & =  (1-u^2)^{m}  \det\left(I_n - u X (I_{2m}+uJ)^{-1} Y^t \right) \notag \\
    & = (1-u^2)^{m-n} \det \left(I_n   -u A+ u^2 Q \right), \notag
\end{align}
where the second line ensues from the generalized matrix determinant lemma and $\det (I_{2m}+uJ) = (1-u^2)^m$. The third line employs the identity $(I_{2m} + uJ)^{-1} = (1-u^2)^{-1}(I_{2m}-uJ) $, while noting $A = XY^t$ and $Q = XJY^t - I_n$. This completes the proof of Eq.~(\ref{eq:bass}). The critical component of this proof involves the use of the generalized matrix determinant lemma, predicated on the factorability of $S$, which can be reinterpreted as an index theorem over a chain complex \cite{hoffman2003remarks}.

Consider a more generalized setup involving an arbitrary set of weights $\u = \{ u_e| e\in E\}$ assigned to each edge. By employing a similar approach, we obtain
\begin{align}\label{eq:bass1}
    \zeta_B(\u)^{-1} = \prod_{e \in E} (1-u_e^2) \det \left(I_n - \tilde A (\u) +\tilde D(\u) \right),
\end{align}
where $\tilde A$ represents the weighted adjacency matrix defined as $\tilde A_{v,v'} = \frac{u_{vv'}}{1-u_{vv'}^2}$, and $\tilde D$ denotes the weighted degree defined as $D_{v,v} = \sum_{(v,v')\in E} \frac{u_{vv'}^2}{1-u_{vv'}^2}$. Specifically, for the $\pm J$ disorders, i.e., $u_{e} = u \tau_{e}$ with $\tau_{e} = \pm 1$, Equation~(\ref{eq:bass1}) simplifies to
\begin{align}
    \zeta_B(u)^{-1} = (1-u^2)^{m-n}\det (I_n - u A'  + u^2 Q ), \notag
\end{align}
where $A'$ includes entries of $0$ and $\pm 1$ to account for bond disorders. It becomes evident that $A'+I$ serves as the adjacency matrix of the percolation model, thereby mapping the Ihara zeta function with $\pm J$ disorder to a percolation problem.

{\bf Manifestly dual formula:}
Consider the fermionic case represented by Eq.~(\ref{eq:Fzeta}), where the winding number is suitably defined only after immersion into a surface with a given spin structure. This marks a notable distinction from its bosonic counterpart in Eq.~(\ref{eq:Bzeta}) and creates a host of technical challenges when applying a similar approach to the one used for the bosonic zeta function. For ease of discussion, our examination is confined to a planar graph $G$ embedded in a plane. However, extending this discussion to surfaces with a higher genus is straightforward.

To explicitly reveal the KW duality, we embed the dual graph $G^*$ over $G$. In this arrangement, each vertex of $G$ is located inside a face of $G^*$, and vice versa; each edge in $G$ intersects with the corresponding edge in $G^*$. For technical convenience, we require these intersections to be perpendicular. Note that the embedding need not be isoradial in general. Figure~\ref{fig:embed}a illustrates such an embedding, where red and blue colors represent $G$ and $G^*$, respectively. 

A crucial element in our depiction involves the quadrilaterals (gray domain in Fig.~\ref{fig:embed}a). Each quadrilateral is formed by two neighboring edges in both $G$ and $G^*$, along with a vertex pair $(v,v^*)$ where $v \in V$ and $v^* \in V^*$. We denote the angles associated with the left and right edges of $v$ and $v^*$ as $\theta_L$ and $\theta_R$, and $\theta_L^*$ and $\theta_R^*$ respectively. These angles satisfy the relations $\theta_L + \theta_R^* = \theta_R + \theta_L^* = \pi/2$ as depicted in Fig.~\ref{fig:embed}a. The collection of $2m$ quadrilaterals, which collectively tile the entire plane, plays a critical role in our new formulation.

\begin{figure}
  \includegraphics[width=1\linewidth]{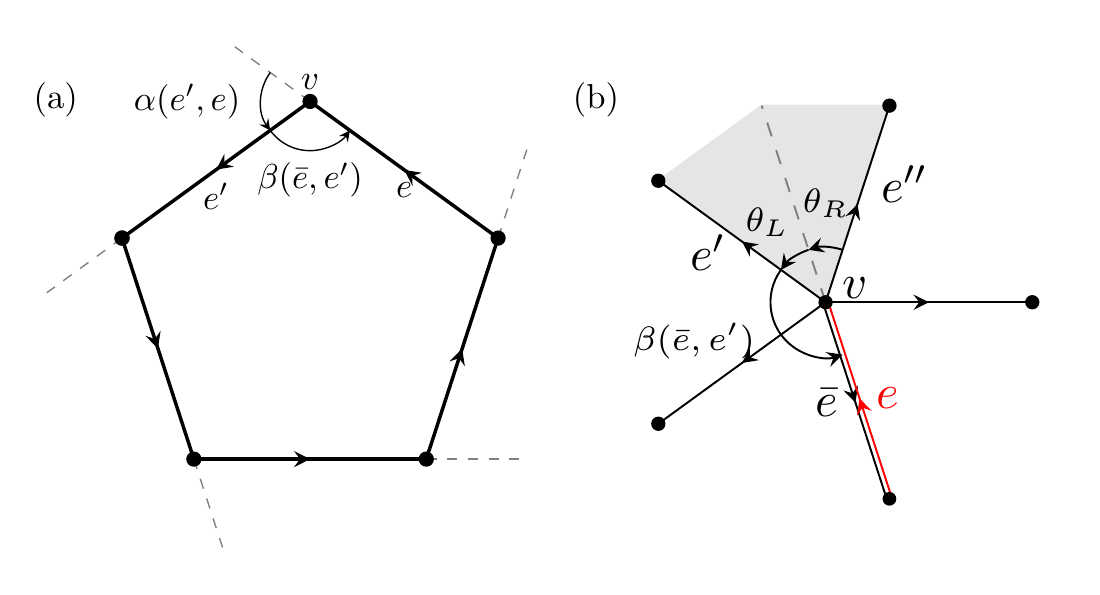}
  \caption{(a) The exterior angle $\alpha(e',e)$ and the interior angle $\beta(\bar e, e')$ for the KW operator satisfy $\alpha(e',e) = \pi - \beta(\bar e, e')$. (b) The angles between two neighboring edges attached to a quadrilateral satisfy $\beta(\bar e,e') = \beta(\bar e,e'') - \theta_{e'} - \theta_{e''}$.
}
  \label{fig:angle}
\end{figure}

The Kac–Ward operator $T_{KW}$ in Eq.~(\ref{eq:Fedge}) is defined similarly to Hashimoto's edge adjacency operator $T$. However, we must introduce a phase change between two consecutive edges to account for the fermionic nature. Specifically, we impose a gauge transformation
\begin{align}
    (T_{KW})_{e',e} = e^{i\alpha(e',e)/2} = i e^{-i \beta(\bar e,e')/2} \notag
\end{align}
if edge $e'$ follows $e$ without backtracking, where $\alpha(e',e)$ is the exterior angle from $e$ to $e'$, and $\beta(\bar e, e')$ is the interior angle between $e'$ and its inverse $\bar e$ (Fig.~\ref{fig:angle}a). This condition ensures that the summation of half exterior angles contributes to a total $\pi$ phase change over a cycle, effectively capturing the fermionic sign in Eq.~(\ref{eq:Fzeta}).

Following a similar approach as that applied to the bosonic zeta function, we introduce the gauged successor operator $S' = T_{KW} - i J $, appending an additional element between edge $e$ and its inverse $\bar e$ with weight $ie^{-i\pi} = -i$. However, $S'$ no longer exhibits factorability, preventing from directly applying the matrix determinant lemma in this scenario. To tackle this issue, we follow the strategy outlined in Refs.~\cite{cimasoni2012critical,cimasoni2015kac}, introducing operator $Q$ between neighboring edges $e$ and $e'$ that share a common starting vertex $v$. Specifically, for each quadrilateral $q$, the operator $Q$ maps its right edge $e$ to the left edge $e'$ while inducing a phase shift $Q_{e,e'} = e^{i(\theta_{L}+\theta_{R})/2}$. As illustrated in Fig.~\ref{fig:angle}b,  $Q_{e',e''} e^{-i\beta(e'',e)/2}  = e^{-i\beta(e',e)/2}$ if $e' \neq \bar e$. However, the fermionic nature possesses a non-trivial monodromy, resulting in a branch cut after a rotation of $2\pi$. This observation leads to the discontinuity
\begin{align}\label{eq:R}
S' - QS' = -2iJ.
\end{align}

As the operator $Q$ acts on the edges attached to the quadrilaterals, it facilitates a natural factorization $Q = L^t R$, where $L$ and $R$ are $2m\times 2m$ matrices associating each quadrilateral $q$ with its left and right edges, respectively. Moreover, we assign weights $L_{q,e} = e^{i\theta_L(q)/2}$ and $R_{q,e} = e^{i\theta_R(q)/2}$. Building on this factorization, we find
\begin{align}
&\det(I-Q)\det(I-uT_{KW}) = \det(I+iuJ -  Q(I-iuJ)) \notag\\
& = (1+u^2)^m\det(I - R  (I-iuJ) (I+iuJ)^{-1} L^t) \notag\\
& = (1+u^2)^m \det\left( I - \frac{1-u^2}{1+u^2} R L^t -  \frac{2u}{1+u^2}  Re^{-i\pi/2}JL^t \right), \notag
\end{align}
where the first equality follows from Eq.~(\ref{eq:R}), and the second stems from the generalized matrix determinant lemma and $\det(I+iuJ) = (1+u^2)^m$. We then introduce the discrete curl operator $d = R L^t$, acting on the space of quadrilaterals, with elements
\begin{align}
    d_{q',q} = e^{i (\theta_L(q) + \theta_R(q'))/2 },\notag
\end{align}
when the quadrilaterals $q'$ and $q$ share a common edge with $q'$ positioned counterclockwise next to $q$. This definition suggests that $d$ can be decomposed into a set of operators $d_v$, each acting on the quadrilaterals around vertex $v$. Therefore, $d = \sum_{v \in V} d_v$, as depicted in Fig.~\ref{fig:embed}b. Similarly, the dual operator $ d_* = \sum_{v^* \in V^*} d_{v^*}$ is the summation of the operators $d_{v^*}$ around the dual vertex $v^*$. 
Observe that $(d_*^\dag)_{q',q} = e^{-i(\pi/2-(\theta_L(q) + \theta_R(q'))/2}$ holds if $q$ possesses a left edge $e$ and $q'$ a right edge $\bar e$ (see Fig.\ref{fig:embed}). This yields $d_*^\dag = Re^{-i\pi/2}JL^t $, which applies to quadrilaterals around dual vertices clockwise. Taking together, we obtain 
\begin{align}\label{eq:mine}
    \zeta_F^{-1} = 2^{-n} (1+u^2)^m \det\left(I_{2m} - \frac{1-u^2}{1+u^2}  d- \frac{1-{u^*}^2}{1+{u^*}^2}  d_*^\dag \right), 
\end{align}
where we employ the identities $\det(I-W) = (1-(-1))^n = 2^n$ and $\frac{1-{u^*}^2}{1+{u^*}^2} = \frac{2u}{1+u^2}$.

Drawing parallels with Eq.~(\ref{eq:bass1}), we can generalize Eq.~(\ref{eq:mine}) to incorporate a set of bond weights $\u$ as follows:
\begin{align}\label{eq:mine1}
    \zeta_F^{-1}(G,\u) = 2^{-n} \prod_{e\in E}(1+u_e^2) \det\left(I_{2m} - D(\u)- D_*^\dag(\u^*) \right)
\end{align}
where $D = R  \frac{1-\u^2}{1+\u^2} L^t$ and $D_*$ represent the weighted curl operators around the vertices in $V$ and $V^*$, respectively. These operators permit a natural decomposition
\begin{align}
    D(\u) = \sum_v D_v{\u}, \ \ \ D(\u^*) = \sum_{v^*} D_{v^*}(\u^*).
\end{align}
The determinant present in Eq.~(\ref{eq:mine1}) is manifestly symmetric under the dual transformation, thereby reinstating the KW duality for an arbitrary set of bond weights \cite{cimasoni2012critical,cimasoni2015kac}
\begin{equation}
    2^{-|V|}\prod_{e\in E} (1+u_e) \zeta_F(G, \u)  = 2^{-|V^*|}\prod_{e \in E} (1+u_e^*) \zeta_F(G^*, \u^*).  
\end{equation}

{\bf Order and disorder operators:} 
Equation~(\ref{eq:mine1}) suggests that the fermionic zeta function $\zeta^{-1}_F = \det(I-H(\u))$ constitutes the characteristic polynomial of a non-Hermitian Hamiltonian:
\begin{align}
    H \equiv D(\u) + D^\dag(\u^*) = \sum_{v\in V} D_{v}(\u) +
 \sum_{v^* \in V^*} D_{v^*}^\dag(\u^*). \notag
\end{align}
Assuming the $\pm J$ disorders of $u_{e} = u \tau_{e}$, where $\tau_{e} = \pm 1$, the equation simplifies to:
\begin{align}\label{eq:RBIM}
    H = \frac{1-u^2}{1+u^2} \sum_{v\in V} d_{v} + \frac{2u}{1+u^2} 
 \sum_{v^* \in V^*} \tilde d_{v^*}^\dag,
\end{align}
where  $(\tilde d_{v^*})_{q,q'} =  e^{i (\theta_L(q) + \theta_R(q'))/2 }\tau_{e}$, applicable for two quadrilaterals sharing the dual vertex $v^*$ and edge $e$. Evidently, only the dual curl operator $D_*$ represents the disorder, while the curl operator $D$ remains unaffected by the disorder. Thus, we interpret $D_v$ and $D_{v^*}$ as local order and disorder operators respectively, echoing the nonlocal disorder operator introduced by Kananoff and Ceva \cite{kadanoff1971determination}.

To establish a connection explicitly, consider a defect that changes a line $\mathcal{L}$ of ferromagnetic bonds to antiferromagnetic, i.e., $\tau_e = -1$ only for $e \in \mathcal{L}$. The corresponding correlation function of nonlocal disorder operators involves a shift in the free energy:
\begin{align}\label{eq:green}
   \Delta F = -\frac{kT}{2} \ln \det(I -  H' G),
\end{align}
where $G = (I-H_0)^{-1}$ represents the Green's function of the ferromagnetic Hamiltonian $H_0 = \frac{1-u^2}{1+u^2}  d- \frac{1-{u^*}^2}{1+{u^*}^2}  d_*^\dag $, and the defect operator $ H' = \frac{2u}{1+u^2} \sum_{v^* \in \mathcal{L}} (\tilde d_{v^*} - d_{v^*})$. It becomes apparent that nonlocal disorder operators correspond to a line integral over the local disorder operator $D_{v^*}$. Consequently, the KW duality presented in Eq.~(\ref{eq:mine1}) reflects an exact interchange of local order and disorder operators under the duality transformation. 

{\bf Implication to RBIM:} 
We now turn to the implications of our new formula for the RBIM. For the sake of technical convenience, our discussion will primarily focus on $\pm J$ disorder on a square lattice, i.e., $P(\tau) = (1-p)\delta(\tau-1) + p \delta(\tau+1)$. A straightforward generalization applies to arbitrary disorder. As we demonstrated earlier, the free energy of $\pm J$ bond disorder is dictated by the spectrum of Eq.~(\ref{eq:RBIM}), where only the disorder operator $\tilde d_*$ accounts for the bond disorders $\tau_e$.

In the absence of disorder ($p = 1$), it is straightforward to diagonalize $H$ via the Fourier transform, which consequently recovers Onsager's free energy. Considering a scenario where $p$ is close to $1$, we can apply the Dyson series expansion to Eq.~(\ref{eq:green}). This approach enables us to determine the critical coupling $u_c(p)$ as a series expansion of disorder probability $1-p$. At the leading order, we find
\begin{equation}
u_c(p) = (\sqrt{2}-1)(1 + 2\sqrt{2} (1-p)) + O((1-p)^2).
\end{equation}
This result aligns with the findings initially obtained using the replica trick \cite{domany1979some}.

We now turn our attention to the zero-temperature limit $\beta \to \infty$ of Eq.~(\ref{eq:RBIM}). In this limit, $(1-u^2)/(1+u^2) \approx 2e^{-2\beta J} $ and $2u/(1+u^2) \approx 1 $. Consequently, in this scenario, disorder operators dominate the spectrum. Direct computation yields
\begin{align}
    \det(I-d_{v^*}^\dag) = 1+W_{v^*},
\end{align}
where the frustration $W_{v^*} \equiv \prod_{e\in P(v^*)}\tau_e = \pm 1$ is defined as the product of edge disorders around the plaquette of the corresponding dual vertices $v^*$. It is clear that when the plaquette is frustrated, i.e., $W_{v^*} = -1$, the determinant acquires a correction from the order operator $d$ with an order at most $e^{-2\beta J}$. On average, there is a $\frac{1-(2p-1)^4}{2}$ chance of $W_{v^*} = -1$, which results in a lower bound of the ground state energy density:
\begin{equation}
    e/J \geq -2+ \frac{1-(2p-1)^4}{2}.
\end{equation}
This finding is consistent with the results in Refs~\cite{vannimenus1977theory,grinstein1979ising} obtained using geometric approaches. 

{\bf Discussion:}
In conclusion, we have unveiled a novel combinatorial approach to Ising models with arbitrary bond weights. In contrast to previous methods, our new formulation distinctly manifests the KW duality via order and disorder operators. We have presented preliminary implications for RBIM at the leading order, and our findings are consistent with results derived from alternative approaches. However, our method has the distinct advantage of seamlessly integrating with the standard framework of perturbative techniques, thereby simplifying the extension to higher-order terms. This also opens up the potential to employ non-perturbative methodologies for a more nuanced understanding of the phase diagram of RBIM. Additionally, our approach can be directly applied to other planar graphs, such as triangular and hexagonal lattices. On the other hand,  it has been suggested that the RBIM may exhibit a disorder duality based on the replica argument, potentially localizing the multicritical point~\cite{nishimori2002duality}. The exactness of this duality and its connection to our method remain unclear. We aim to explore these questions in future research.

Our methodology can also be readily generalized to anyonic statistics by considering a non-half-integer phase shift, a topic we plan to discuss elsewhere. Further, a non-Abelian generalization seems feasible. These generalizations have close ties with parafermionic models \cite{fradkin1980disorder}. Moreover, given that the Ihara zeta function can generalize to high-dimensional objects \cite{kang2014zeta}, it is enticing to contemplate a similar higher-dimensional generalization for its fermionic counterpart. Such an extension may hold promise for a solution to the 3D Ising model.

%\begin{acknowledgments}
%We express our gratitude to Aninda Sinha and Ahmadullah Zahed for their invaluable discussions and constructive feedback on the manuscript.
%\end{acknowledgments}

%\bibliographystyle{unsrt}
\bibliography{ref}% Produces the bibliography via BibTeX.

\end{document}